\documentclass[aip,amsmath,amssymb,reprint]{revtex4-1}

\pdfoutput=1 

\usepackage{graphicx}
\usepackage{dcolumn}
\usepackage{bm}

\usepackage[utf8]{inputenc}
\usepackage[T1]{fontenc}
\usepackage{mathptmx}
\usepackage{etoolbox}

\usepackage{amsbsy}
\usepackage{colordvi}
\usepackage{color}     
\usepackage{latexsym}
\usepackage{tikz}
\usetikzlibrary{positioning,patterns}

\makeatletter
\def\@email#1#2{%
 \endgroup
 \patchcmd{\titleblock@produce}
  {\frontmatter@RRAPformat}
  {\frontmatter@RRAPformat{\produce@RRAP{*#1\href{mailto:#2}{#2}}}\frontmatter@RRAPformat}
  {}{}
}%
\makeatother

\newcommand{\bB}{\mathbf{B}}			
\newcommand{\dd}{\mathrm{d}}			
\newcommand{\bF}{\mathbf{f}}			
\newcommand{\Demag}{\bar{\mathbf{D}}}		
\newcommand{\bh}{\mathbf{h}}			
\newcommand{\bH}{\mathbf{H}}			
\newcommand{\bM}{\mathbf{M}}			
\newcommand{\br}{\mathbf{r}}			
\newcommand{\bR}{\mathbf{R}}			
\newcommand{\bu}{\mathbf{u}}			
\newcommand{\bv}{\mathbf{v}}			
\newcommand{\bvs}{\mathbf{v}'}
\newcommand{\bV}{\mathbf{V}}
\newcommand{\bW}{\mathbf{W}}			
\newcommand{\bvpec}{\tilde{\mathbf{v}}}	
\newcommand{\tauB}{\tau_{\rm B}}		
\newcommand{\tB}{t_{\rm B}}		
\newcommand{\bomega}{\boldsymbol{\omega}}
\newcommand{\bOmega}{\boldsymbol{\Omega}}
\newcommand{\bnabla}{\boldsymbol{\nabla}}
\newcommand{\kT}{k_{\rm B}T}			
\newcommand{\chiL}{\chi_{\rm L}}		
\newcommand{\bnormal}{\hat{\mathbf{k}}}
\newcommand{\betath}{\beta_{\rm th}}	 
\newcommand{\MM}{Q}					

\newcommand{\ave}[1]{\langle #1 \rangle}

\newcommand{\changed}[1]{{#1}} 

\begin{document}

\title{Multiparticle Collision Dynamics for Ferrofluids}

\author{Patrick Ilg}

\affiliation{School of Mathematical, Physical, and Computational Sciences, University of Reading, Reading, RG6 6AX, United Kingdom}
\email{p.ilg@reading.ac.uk}

\date{\today}

\begin{abstract}
Detailed studies of the intriguing field-dependent dynamics and transport properties of confined flowing ferrofluids require efficient mesoscopic simulation methods that account for fluctuating ferrohydrodynamics. 
Here, we propose such a new mesoscopic model for the dynamics and flow of ferrofluids, where we couple the multi-particle collision dynamics method as a solver for the fluctuating hydrodynamics equations to the stochastic magnetization dynamics of suspended magnetic nanoparticles. 
This hybrid model is validated by reproducing the magnetoviscous effect in Poiseuille flow, obtaining the rotational viscosity in quantitative agreement with theoretical predictions. 
We also illustrate the new method for the benchmark problem of flow around a square cylinder. 
Interestingly, we observe that the length of the recirculation region is increased whereas the drag coefficient is decreased in ferrofluids when an external magnetic field is applied, compared with the field-free case at the same effective Reynolds number. 
The presence of thermal fluctuations and the flexibility of this particle-based mesoscopic method provides a promising tool to investigate a broad range of flow phenomena of magnetic fluids and could also serve as an efficient way to simulate solvent effects when colloidal particles are immersed in ferrofluids. 
\end{abstract}

\maketitle

\section{Introduction}

Colloidal suspensions of magnetic nanoparticles, also known as ferrofluids, are fascinating model systems that combine superparamagnetic and magnetoviscous effects \cite{rosensweigbook,Odenbach_LNP763}.
The classical experiments of McTague \cite{mctague} demonstrated that pipe flow of ferrofluids can be manipulated by external magnetic fields, showing anisotropic response depending on the relative orientation of the magnetic field with respect to the flow direction. 
Shliomis and co-workers \cite{MRS74} successfully explained these experiments by a kinetic theory of hindered particle rotations and the resulting changes in rotational viscosity. 
These ground-breaking works sparked numerous subsequent studies on the magnetoviscous effect, leading to more refined experiments and a better theoretical understanding, as well as several novel applications of ferrofluids that rely on their field-dependent flow behaviour (see e.g.\ \cite{ODENB02,KrogerIlgHess_JPHYS03,Parak_review,Ilg_lnp,felicia_recent_2016} and references therein). 
Also various computer simulation methods for ferrofluid flow have been developed in recent years. 
The simulations range from simple channel or pipe flows \cite{Rinaldi_spin,Finlayson2003,Hirabayashi:2001hb} 
to more complicated geometries \cite{Sheikholeslami:2018dp} including free-surface flows \cite{LiboHuang:2019de}. 
Different numerical methods were used in these studies, ranging from perturbative solutions \cite{Rinaldi_spin} to adaption of finite-element and finite-volume computational fluid dynamics codes \cite{Finlayson2003,Sheikholeslami:2018dp}, the Lattice Boltzmann method \cite{Hirabayashi:2001hb}, and smooth particle hydrodynamics simulations \cite{LiboHuang:2019de}. 
Such flow simulations are helpful for various applications, e.g.\ planning magnetic drug targeting treatments \cite{kayal_flow_2011}. 
Also micro magnetofluidics shows promising potential for contactless mixing, separation, and trapping of particles and polymers on small scales, where flow simulations help to improve their effectiveness \cite{Munaz:2018cw}. 

Thermal fluctuations are important in small-scale flows which, however, are neither captured by traditional computational fluid dynamics codes, nor in the Lattice Boltzmann method. 
Instead, multi-particle collision (MPC) dynamics provides a very versatile and flexible method for simulating fluid flow including thermal fluctuations \cite{malevanets_mesoscopic_1999,Ihle:2001bf,Gompper:2009is}. 
In this off-lattice method, each particle represents a small fluid element and evolves through a sequence of streaming and simplified collision steps. By locally conserving mass, momentum and energy, the correct hydrodynamic behavior is generated. In addition, thermal fluctuations are naturally present in this particle-based scheme. 
Recently, the MPC method has been extended to model anisotropic fluids like nematic liquid crystals \cite{Shendruk:2015jw,Lee:2015iv,Mandal:2019bs}. 
Here, we use similar reasoning to extend the MPC method to model ferrofluid flow including stochastic magnetization dynamics. 
We note that none of the previous approaches to ferrofluid flow simulations \cite{Rinaldi_spin,Finlayson2003,Sheikholeslami:2018dp,Hirabayashi:2001hb,LiboHuang:2019de} includes thermal fluctuations in the magnetization dynamics.  
For simplicity, we restrict ourselves to a quasi-two-dimensional system, i.e.\ two-dimensional translational motion with three-dimensional magnetization dynamics. 
We demonstrate that this model is able to reproduce fluctuating ferrohydrodynamics. 
In particular, we verify that the field-dependent effective viscosity resulting from this hybrid method quantitatively agrees with theoretical predictions. 
Furthermore, we show the flexibility of this approach by simulating the flow around a square cylinder. 
Analyzing the flow field, we determine the length of the recirculation region and find it decreasing with increasing field strength. 
Moreover, we also calculate the drag coefficient and study its dependence on the magnetic field. 

The new hybrid method presented here is a useful tool that also allows to study the effect of fluctuations on ferrofluid flow on a mesoscopic scale.   Different models of magnetization dynamics can be incorporated in a straightforward manner. Thereby, the flexibility of the MPC method also facilitates the study of more complicated geometries. 
This paper is organized as follows. 
In Sect.\ \ref{ferrohydro.sec}, we briefly review the ferrohydrodynamic theory.  
The hybrid MPC model coupled to stochastic magnetization dynamics is 
described in Sect.\ \ref{model.sec}. 
Sect.\ \ref{simu.sec} describes the simulation setup, before results are presented and discussed for channel flows in Sect.\ \ref{results-channel.sec} and for flow around square cylinder in Sect.\ \ref{results-cylinder.sec}. 
Finally some conclusions are offered in Sect.\ \ref{conclude.sec}. 

\section{Ferrohydrodynamics} \label{ferrohydro.sec}

We here give a brief overview of the basic equations of  ferrohydrodynamics to make the paper self-contained. Further details can be found e.g.\ in Refs.\ \cite{rosensweigbook,roseninlnp,Shlio_LNP}. 
The fluid momentum balance equation reads 
\begin{equation} \label{momentum-eq}
\frac{\dd\bv}{\dd t}   = 
-\bnabla p/\rho + \nu \nabla^{2} \bv  +  \bF_{\bM}
\end{equation}
with $\bv$ the fluid velocity field, $\dd\bv/\dd t$ denotes the material derivative, 
$\rho$ and $\nu$ are the fluid density and kinematic viscosity, respectively, 
and $p$ the scalar pressure. 
With allowance for internal rotations, the influence of magnetization effects on fluid flow is described by the force density 
\begin{equation} \label{Fmag}
\rho \bF_{\bM} = (\bM\cdot\bnabla)\bH + \frac{1}{2}\bnabla\times(\bM\times\bH)
\end{equation}
where the magnetic field and the magnetization are denoted by $\bH$ and $\bM$, respectively. 
We assume the fluid to be incompressible and non-conducting, therefore \begin{equation} \label{magnetostatics.eq}
\bnabla\cdot\bv = 0, \quad
\bnabla\times \bH = {\bf 0}, \quad
\bnabla\cdot{\bf B} = 0, 
\end{equation}
where ${\bf B}=\mu_{0}(\bH+\bM)$ denotes the magnetic induction and $\mu_0$ the permeability of free space. 

Equations \eqref{momentum-eq} -- \eqref{magnetostatics.eq} are not closed due to the appearance of the magnetization $\bM$. 
Some of the previous simulation approaches have assumed quasi-equilibrium conditions and locally approximate $\bM\approx \bM_{\rm eq}(\bH)$ \cite{Sheikholeslami:2018dp,LiboHuang:2019de}. 
Such approaches, however, neglect relaxation phenomena and a more accurate treatment of the magnetization dynamics is often desirable. 
Unfortunately, despite many efforts and long debates in the literature about the appropriate form of the magnetization equation for ferrofluids, no consensus has been reached yet (see e.g.\ Refs.\ \cite{Shlio_LNP,Luecke_ZChem,Liu_LNP,roseninlnp,Ilg_lnp} 
and references therein). 

One of the advantages of the present method is that it can handle different magnetization equations rather straightforwardly. 
Since the scheme naturally includes thermal fluctuations in the hydrodynamic variables, we here choose to implement a mesoscopic model that also accounts for thermal fluctuations in the magnetization. 
In particular, we employ the classical model of ferrofluid dynamics of rigid dipoles proposed in Ref.\ \cite{MRS74} and widely studied since \cite{Ilg_lnp,IKH01,soto-aquino_oscillatory_2011}. %
In the free-draining limit and neglecting inter-particle interactions, the torque balance of viscous, magnetic, and Brownian torques reads  
\begin{equation} \label{torque-balance}
-\zeta (\bomega_{i} - \bOmega(\br_i)) + \mu \bu_{i} \times \bH(\br_i) + {\bf T}^{\rm B}_i = {\bf 0},
\end{equation}
where $\mu\bu_{i}$ and $\bomega_{i}$ are the magnetic moment and angular velocity of magnetic nanoparticle $i$, respectively. 
The magnitude of the magnetic moment is denoted by $\mu$. 
The rotational viscous torque is proportional to the rotational friction coefficient $\zeta$ and arises when the particle rotation does not match the local vorticity, $\bOmega = (1/2)\bnabla\times\bv$.  
The magnetic torque is generated by the magnetic field $\bH$ at the location of the nanoparticle. 
Brownian fluctuations are modelled by random torques  
${\bf T}^{\rm B}_i$ with Gaussian white noise, 
$\ave{{\bf T}^{\rm B}_i}=0$ and  
$\ave{{\bf T}^{\rm B}_i(t){\bf T}^{\rm B}_j(t')}=2\kT\zeta \delta_{ij}\delta(t-t') {\bf I}$, 
$k_{\rm B}$ and $T$ Boltzmann's constant and temperature, respectively, and 
${\bf I}$ the three-dimensional unit matrix. 
The fluid magnetization is calculated from local averages over the magnetic moments of the nanoparticles as 
$\bM(\br,t) = \ave{\sum_{i} \mu \bu_{i}(t)w(\br - \br_{i}(t))}$, where $w$ is a suitable weight function. 
The magnetization arising from an applied field is responsible for the backflow effect via the magnetic force density given in Eq.\ \eqref{Fmag}. 
\changed{In equilibrium, the model describes superparamagnetic behavior, $\bM_{\rm eq}\sim L_1(H)\bH/|\bH|$, with $L_1$ the Langevin function \cite{rosensweigbook}.}

\section{Model and Numerical implementation} \label{model.sec}

MPC is a very versatile and flexible, particle-based method to simulate fluid flow. 
The method has recently been extended to complex fluids \cite{Gompper:2009is} like 
\changed{polymer solutions \cite{kowalik_multiparticle_2013}} and in particular also to anisotropic fluids like nematic liquid crystals 
\cite{Shendruk:2015jw,Lee:2015iv,Mandal:2019bs}. 
Inspired by these recent works, we here present an application to another class of anisotropic fluids, namely ferrofluids. 

Within MPC, the fluid is represented by a system of $N$ identical particles, each with mass $m_i=m$, where $\br_{i}$ and $\bv_{i}$ denote the position and velocity of particle $i$ and $i=1,\ldots,N$. 
Here, \changed{we follow previous works \cite{kowalik_multiparticle_2013,Shendruk:2015jw,Lee:2015iv,Mandal:2019bs} and let an MPC particle represent a coarse-grained description of a fluid element, containing solvent as well as solute particles.}
\changed{Thus,} each \changed{MPC} particle also carries a magnetic moment $\mu\bu_{i}$, where $\mu$ denotes the magnitude and the three-dimensional unit vector $\bu_{i}$  describes the orientation of the magnetic moment of particle $i$. 
The basic idea underlying MPC is that particle dynamics can be split into a streaming and a simplified collision step, both can be computed very efficiently. 
For anisotropic fluids, the translational motion needs to be coupled to rotational dynamics as detailed below for the case of ferrofluids. 

In the streaming step, the positions and velocities of the particles are updated according to \cite{JonathanKWhitmer:2010kp}
\begin{align} \label{streaming.eq}
\br_{i}(t+\Delta t)  & = \br_{i}(t) + \Delta t\, \bv_{i}(t) + \frac{\Delta t^{2}}{2m_i}\,{\bf F}_{i}(t)\\
\bvs_{i}(t)& = \bv_{i}(t) + \frac{\Delta t}{m_i}\, {\bf F}_{i}(t)
\end{align}
with time step $\Delta t$. 
The force ${\bf F}_{i}=\bF_{\rm ext} + \bF_{\bM}(\br_{i})$ acting on particle $i$ consists of a uniform external forcing $\bF_{\rm ext}$ (e.g.\ due to an applied pressure gradient) and the magnetic force $\bF_{\bM}$ defined in Eq.\ \eqref{Fmag}. 

The collision step ensures momentum exchange between particles and can be expressed as \cite{Lee:2015iv}
\begin{equation} \label{collision.eq}
\bv_{i}(t+\Delta t) = \bV_{\!C_{i}}(t) +  \betath\bR\cdot[\bvs_{i}(t) - \bV_{\!C_{i}}(t)]
\end{equation}
where 
$\bV_{\!C_{i}}$ is the center-of-mass velocity of the collision cell to which particle $i$ belongs, 
$\bV_{\!C_{i}}=\sum_{j\in {\cal C}_{i}}m_{j}\bvs_{j}/M_{C_{i}}$. 
The total mass in this collision cell is given by  
$M_{C_i}=\sum_{j\in{\cal C}_i}m_j$, 
and ${\cal C}_{i}$ labels all particles currently located in collision cell $C_{i}$. 
In MPC, the collision step \eqref{collision.eq} is performed simultaneously between all particles $j$ currently residing in the same collision cell. These cells are defined by a regular square grid of length $a$. 
Collisions lead to rotation of the relative velocities, which is described by the rotation matrix ${\bR}$ in Eq.\ \eqref{collision.eq}. In two spatial dimensions considered here, ${\bf R}$ is completely specified by the rotational angle $\alpha$. Thus, ${\bf R}$ rotates the relative velocities by an angle $\pm\alpha$ with equal probabilities. 

As has been emphasized before \cite{Gompper:2009is}, angular momentum conservation is important for the rotational dynamics of anisotropic fluids, but is in general violated by Eq.\ \eqref{collision.eq}. 
Angular momentum conservation can be restored for this scheme, however, if the rotation angle $\alpha$ is not fixed, but chosen in collision cell $C_i$ according to 
\begin{equation}
	\cos \alpha = \frac{A_1^2-A_2^2}{A_1^2 + A_2^2}, \quad 
	\sin \alpha = -\frac{2A_1A_2}{A_1^2 + A_2^2}
\end{equation}
where $A_1=\sum_{j\in{\cal C}_i}[\br_j\times\bvpec_j]_z$, 
$A_2=\sum_{j\in{\cal C}_i}\br_j \cdot \bvpec_j$, with  
relative velocities $\bvpec_{i}=\bvs_{i} - \bV_{\!C_{i}}$ \cite{Gompper:2009is}. 

In flow simulations, a thermostat is generally needed in order to remove the energy input into the system. Here, we follow common practice (see e.g.\ Ref.\ \cite{Lee:2015iv}) and use a simple rescaling of the relative velocities to the bath temperature $T$ by $\betath=\sqrt{T/T_{C_{i}}}$. 
The instantaneous kinetic temperature in collision cell $C_{i}$ is defined in two spatial dimensions by 
$\kT_{C_{i}}=(1/2N_{C_i})\sum_{j\in {\cal C}_{i}}m_{j}\bvpec_{j}^{2}$, with $N_{C_i}=\sum_{j\in{\cal C}_i}1$ the number of particles in the collision cell. 

Using a fixed grid of collision cells violates Galilean invariance and might lead to spurious correlations. Therefore, following Ref.\ \cite{Ihle:2001bf}, in each time step, the grid is shifted by a random vector, where each component is uniformly distributed in $[-a/2,a/2]$.

The simulations discussed below employ two types of boundary conditions. 
Bulk simulations employ periodic boundary conditions in all spatial dimensions. For flow in a channel geometry, we employ periodic boundary conditions only in the flow direction, while no-slip boundary conditions on the confining walls are realized by the well-known bounce back rule, where the velocity of the particles are reversed as a result of collision with the wall \cite{JonathanKWhitmer:2010kp}. 
Since random grid shifting can lead to underpopulated cells at the walls that would undermine the no-slip condition, additional, so-called ghost particles are added in the collision step \cite{JonathanKWhitmer:2010kp}. 

So far, we have specified the translational motion only. 
To simulate ferrofluid flow, we also need to specify the dynamics of the magnetic moments of the particles. 
From the kinematic equation for rotations, $\dot{\bu}_i=\bomega_i\times\bu_i$, and Eq.\ \eqref{torque-balance}, a weak first-order scheme can be used, 
$\bu_i(t+\Delta \tB)=\bu_i^{\rm P}/|\bu_i^{\rm P}|$, where 
$\bu_i^{\rm P}=\bu_i(t)+\Delta\bomega_i\times\bu_i(t)$ with 
\begin{equation}\label{delta-omega}
	\Delta\bomega_i = [\tauB \bOmega_{C_i} + \frac{1}{2}\bu_i\times\bh_{C_i}]\frac{\Delta \tB}{\tauB} + \frac{1}{\sqrt{\tauB}}\Delta\bW_i
\end{equation}
In Eq.\ \eqref{delta-omega} we 
introduced the dimensionless magnetic field, which is defined as the ratio of the local Zeeman energy over the thermal energy, $\bh_{C_i}=\mu\bH(\br_{C_i})/\kT$. 
Furthermore, $\Delta\bW_i$ denotes the Wiener increment over a time interval $\Delta \tB$ and 
$\tauB=\zeta/(2\kT)$ the Brownian rotational relaxation time of a single nanoparticle. 
Equation \eqref{delta-omega} requires $\Delta \tB/\tauB\ll 1$. 
For the simulations shown below, we use a second-order stochastic Heun algorithm. More details on the algorithm are given in appendix \ref{Heun.sec}. 

In Eq.\ \eqref{delta-omega}, we denote the local vorticity and local magnetic field with $\bOmega_{C_i}$ and $\bh_{C_i}$, respectively, to indicate that these fields are not evaluated on the precise location of particle $i$, but at the center of the collision cell to which particle $i$ belongs at this time step. 
Similarly, we define the instantaneous magnetization of the collision cell as 
$\bM_{C_i}=(M_{\rm sat}/N_{C_i})\sum_{j\in{\cal C}_i}\bu_j$, with the saturation magnetization $M_{\rm sat}=n\mu$ and $n$ the number density of magnetic nanoparticles. 
For the mean magnetization, we recover the usual relation 
$\ave{\bM_{C_i}}=M_{\rm sat}\ave{\bu}$. 

The magnetization dynamics \eqref{torque-balance} is coupled to the flow field via the vorticity, i.e.\ gradients of the velocity field, $\bOmega=(1/2)\bnabla\times\bv$, whereas gradients in the magnetization or magnetic field lead to backflow effects via the magnetic force density in the momentum balance, Eq.\ \eqref{Fmag}. 
In order to compute these gradients, we evaluate the velocity and magnetization field in each time step via a kernel smoothing method and determine spatial gradients from a finite-difference approximation. Further details on the numerical procedure are provided in appendix \ref{kernelsmooth.sec}.

\section{Simulation setup} \label{simu.sec}

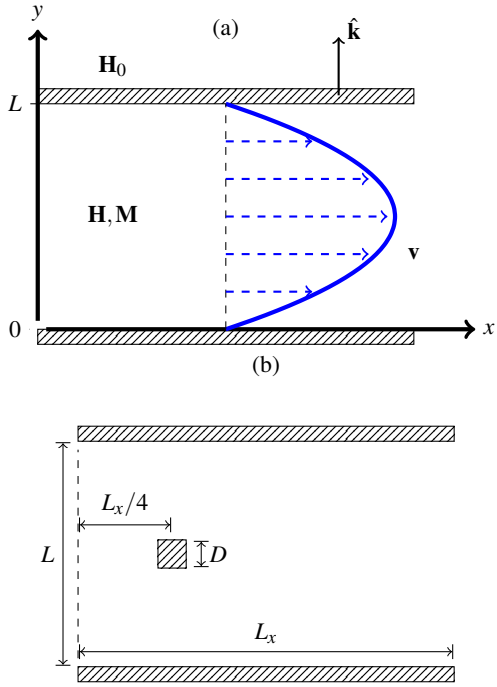
\begin{figure}[h!]
\def\Ly{3}
\def\Lx{5}
\def\xp{2.5}
\begin{tikzpicture}
\pgfmathsetmacro{\Lyzwei}{2*\Ly}
\node at ({\Lx/2},{\Ly+1}) () {(a)};
\node at (0,0) (a) {};
\node at (\Lx,0) (b) {};
\node at ({\Lx+1},0) (bb) {$x$};
\node at (0,\Ly) (c) {};
\node at (0,{\Ly+1.2}) (cc) {$y$};
\node at (-0.3,\Ly) (L) {$L$};
\draw[-] (L) -- (0,\Ly);
\node at (-0.3,0) (O) {$0$};
\draw[-] (O) -- (a);
\node at (\Lx,\Ly) (d) {};
\node at (1,{\Ly/2}) (Hin) {$\bH, \bM$};
\node at (1,3.5) (Hext) {$\bH_{0}$};
\node at (4,\Ly) (k) {};
\node at (4,{\Ly+1}) (kk) {};
\draw[pattern=north east lines] (a) rectangle ([yshift=-0.2cm]b);
\draw[pattern=north east lines] (c) rectangle ([yshift=0.2cm]d);
\draw[->,ultra thick] (a) -- (bb);
\draw[->,ultra thick] (a) -- (cc);
\draw[->,thick] (k) -- (kk) node[right] {$\hat{\bf k}$};
\draw[scale=1, domain=0:3, smooth, variable=\y, blue, ultra thick]  plot ({\xp+\y*(\Ly-\y)}, {\y});
\draw[dashed] (\xp,0) -- (\xp,\Ly); 
\draw[dashed,->,thick,blue] (\xp,0.5) -- (3.65,0.5);
\draw[dashed,->,thick,blue] (\xp,1) -- (4.4,1);
\draw[dashed,->,thick,blue] (\xp,1.5) -- (4.65,1.5);
\draw[dashed,->,thick,blue] (\xp,2) -- (4.4,2);
\draw[dashed,->,thick,blue] (\xp,2.5) -- (3.65,2.5);
\node at (5,1) {$\bv$}; 
\end{tikzpicture}
\begin{tikzpicture}
\pgfmathsetmacro{\D}{\Ly/8}
\pgfmathsetmacro{\xleft}{\Lx/4 - \D/2}
\pgfmathsetmacro{\xright}{\Lx/4 + \D/2}
\pgfmathsetmacro{\ydown}{\Ly/2 - \D/2}
\pgfmathsetmacro{\yup}{\Ly/2 + \D/2}
\node at ({\Lx/2},{\Ly+1}) () {(b)};
\node at (0,0) (a) {};
\node at (-.1,.2) (aa) {};
\node at (-.2,-.1) (aaa) {};
\node at (\Lx,0) (b) {};
\node at ({\Lx+.1},.2) (bb) {};
\node at (0,\Ly) (c) {};
\node at (-.2,{\Ly+.1}) (cc) {};
\node at (\Lx,\Ly) (d) {};
\node at (\xleft,\ydown) (C1) {};
\node at (\xright,\ydown) (C2) {};
\node at ({\xright+.2},{\ydown-.1}) (C2r) {};
\node at (\xright,\yup) (C3) {};
\node at ({\xright+.2},{\yup+.1}) (C3r) {};
\node at (\xleft,\yup) (C4) {};
\draw[pattern=north east lines] (C4) rectangle (C2);
\node at (-.1,{\yup+.2}) (ac) {};
\node at ({\Lx/4+.1},{\yup+.2}) (bc) {};
\draw[pattern=north east lines] (a) rectangle ([yshift=-0.2cm]b);
\draw[pattern=north east lines] (c) rectangle ([yshift=0.2cm]d);

\draw[|<->|] (aa) -- (bb) node[pos=.5,above] {$L_x$};
\draw[|<->|] (ac) -- (bc) node[pos=.5,above] {$L_x/4$};
\draw[|<->|] (aaa) -- (cc) node[pos=.5,left] {$L$};
\draw[|<->|] (C2r) -- (C3r) node[pos=.5,right] {$D$};
\draw[dashed] (a) -- (c);
\end{tikzpicture}

\caption{(a) Schematic of channel geometry and velocity profile. (b) Channel including square cylinder in Sect.~\ref{results-cylinder.sec}.}
\label{channel.fig}
\end{figure}

Consider a simple channel geometry of width $L$ with infinite plates  perpendicular to the $y$-direction, as sketched in Fig.\ \ref{channel.fig}. 
The externally applied field is denoted by $\bH_{0}$. We assume the exterior to be non-magnetic such that $\bM_{0}={\bf 0}$ with the external magnetic induction $\bB_{0}=\mu_0 \bH_{0}$. 
The magnetic field inside the channel is denoted by $\bH$. The presence of ferrofluid gives rise to a magnetization $\bM$, such that the magnetic induction inside the channel is given by $\bB=\mu_0(\bH+\bM)$.  

If $\bnormal$ denotes the normal vector on the interface, the continuity conditions of magnetostatics require 
 the normal component of the magnetic induction $\bB$ and the tangential component of the magnetic field $\bH$ are continuous at the interface
\cite{rosensweigbook}
\begin{equation} \label{field-continuity.eq}
\bnormal \cdot(\bB_{0}-\bB)= 0, \quad
\bnormal\times(\bH_{0}-\bH) = {\bf 0}.
\end{equation}

The internal field can be expressed as $\bH=\bH_{0}-\Demag\cdot\bM$ where the demagnetization tensor $\Demag$ depends on the geometry. 
For the simple channel geometry of Fig.\ \ref{channel.fig} with infinite plates and a spatially homogeneous external field $\bH_0$, we obtain 
$H_x=H_{0,x}$ and $H_y=H_{0,y}-M_y$. 
The same result is obtained for an infinite needle-shaped ellipsoidal geometry \cite{Osborne1945}. 
It is important to note that Maxwell's equation together with the continuity conditions \eqref{field-continuity.eq} require for our case the magnetization $\bM$ to be independent of $x$-position, thus $\bM=\bM(y)$ can depend only on the vertical position within the channel. Further details are provided in Appendix \ref{magnetostatics.sec}. 
In terms of the dimensionless field $\bh$, the relation between external and internal field can be written as 
$h_y=h_{0,y}-3\chiL\ave{u_y}$ with the Langevin susceptibility $\chiL=n\mu^2/(3\kT)$. 
 
In order to study Poiseuille flow, we model an applied pressure gradient as a homogeneous external forcing in $x$-direction, $\bF_{\rm ext}=f_{\rm ext}\hat{\bf i}$. This constant force is applied to all particles, see Eq.\ \eqref{streaming.eq}. 
For the channel geometry of Fig.~\ref{channel.fig}, the dimensionless magnetic force density $\bF_{\bM}^{\ast}=(\Delta t^{2}/a)\bF_{\bM}$ from Eq.\ \eqref{Fmag} becomes  
\begin{align} \label{Fmag-dimless.eq}
\bF_{\bM}^{\ast} & = -n^{\ast}
\left( \begin{array}{c}
\frac{1}{2}[(3\chiL \ave{u_{y}}- h_{0,y})\frac{\dd \ave{u_{x}}}{\dd y^{\ast}} + (h_{0,x}+3\chiL\ave{u_{x}})\frac{\dd \ave{u_{y}}}{\dd y^{\ast}}]\\
3\chiL\ave{u_{y}}\frac{\dd \ave{u_{y}}}{\dd y^{\ast}}
\end{array}\right)
\end{align}
where $n^{\ast}=n\kT/p_{\rm ref}$ is a density ratio and $p_{\rm ref}=\rho a^2/\Delta t^2$ a reference pressure. In Eq.\ \eqref{Fmag-dimless.eq}, we used the fact that the magnetostatic fields are independent of the $x$-coordinate. 

We follow common practice in MPC simulations and choose particle masses as $m_i=1$, the linear size of collision cell $a=1$, and time step $\Delta t=1$. 
\changed{We note that we limit ourselves to small forcings, otherwise smaller values of $\Delta t$ are needed.}
Having set the basic MPC units, table \ref{params.table} gives an overview of the remaining model parameters and the values or range of values used in subsequent simulations.  A number of comments are in order. 
Temperature is measured in units of $T_{\rm ref} = m v_{\rm max}^2/k_{\rm B}$, with $v_{\rm max}=a/\Delta t$ the maximum propagation speed. 
The mean free path is defined as $\lambda=\Delta t\sqrt{\kT/m}$. 
We choose $T^\ast=T/T_{\rm ref}<1$ to ensure the mean free path is smaller than the grid size. 
The mean number of MPC particles per collision cell $\MM$ is chosen between $20$ and $100$. 
Furthermore, we choose $\tauB^\ast = 100$ to model slower relaxation of nanoparticles compared to the base fluid. 
This then allows us to set $\Delta \tB = \Delta t$, satisfying $\Delta\tB/\tauB=10^{-2}\ll 1$. 

Since every MPC particle represents a small volume element of fluid, there is no contradiction of choosing $n^\ast$ independent of $\MM$. 
Although particles in the original MPC scheme are point-like, introducing the rotational friction coefficient $\zeta$ in Eq.\ \eqref{torque-balance} allows us to associate an effective hydrodynamic diameter $\sigma$ to the particles via Stokes' formula $\zeta = \pi\eta_{\rm s}\sigma^3$, where $\eta_{\rm s}$ denotes the viscosity of the solvent. Here, we refer to the pure MPC fluid as ``solvent''. Therefore, it is possible to express the density ratio $n^\ast$ in terms of the volume fraction of magnetic particles $\phi$ as $n^\ast=3\phi\nu_{\rm s}^\ast/\tauB^\ast$, where we used the definition of the reference pressure $p_{\rm ref}$ and the dimensionless kinetic viscosity $\nu_{\rm s}^\ast=\nu_{\rm s}\Delta t/a^2$ with $\nu_{\rm s}=\eta_{\rm s}/\rho$. 
Therefore, even though every particle in the simulation carries a magnetic moment $\mu$, the condition $n^\ast\tauB^\ast/\nu_{\rm s}^\ast\ll 1$ still corresponds to dilute suspensions of magnetic nanoparticles, which, strictly speaking, is needed for justifying the magnetization dynamics described by Eq.\ \eqref{torque-balance}. 

\begin{table}
\centering
\begin{tabular}{lcccccccc}
\hline\hline
  variable \phantom{.} & $\MM$ & $T^\ast$ & \phantom{..}$\tauB^\ast$ \phantom{..} &  \phantom{.}$\chiL$ \phantom{.} & $h$ & $n^\ast$ & $f_{\rm ext}^\ast$ & $v_{\rm max}$\\
  values \phantom{.} & 20--100 & 0.05--0.4 & 100& $0$ & 0--5 & 0--0.005 & 0--$5\!\times\!10^{-5}$ & 0--0.1\\
  \hline\hline
  \end{tabular}
\caption{Overview of model parameters and their numerical values used in these simulations: mean number of MPC particles per collision cell $\MM=\ave{N_{C_i}}$, reduced temperature $T^\ast=T/T_{\rm ref}$, dimensionless Brownian relaxation time $\tauB^\ast=\tauB/\Delta t$, the Langevin susceptibility $\chiL$, the Langevin parameter $h=\mu H/\kT$, the density ratio $n^\ast=n\kT/p_{\rm ref}$, the strength of the external forcing $f_{\rm ext}^\ast$, and the maximum velocity $v_{\rm max}$.}
\label{params.table}	
\end{table}

\section{Results for channel flow} \label{results-channel.sec}

All simulations are started from an equilibrium initial state, where random velocities are assigned to the particles according to the Maxwell-Boltzmann distribution at temperature $T$. Initially, particles are placed at random positions to uniformly fill the simulation box and the orientations $\bu_i$ of the particles' magnetic moments are chosen randomly from the three-dimensional isotropic distribution. 
Typically, simulations are run for $10^5$ integration steps and averages are collected only in the second half of the run where steady-state has been reached. 

For $n^\ast=0$, backflow effects are absent and the standard translational dynamics of the MPC fluid is recovered. 
We tested our numerical implementation for this case against literature data, dropping angular momentum conservation and choosing a fixed rotation angle $\alpha$. 
For equilibrium bulk simulations with periodic boundary conditions in both spatial dimensions, we recover the theoretical result for the self-diffusion coefficient, $D^\ast = T^\ast \Delta t ( 1/2 + b/[1-b] )$ with 
$b = 1/\MM  + (1 - 1/\MM)\cos\alpha$,
where $\MM=\rho a^{2}/m$ is the average number of particles in a collision cell  \cite{Ihle:2001bf}. For this test, we chose $\MM=20$ and varied $\alpha$ between 70 and 120 degrees. 

As a second test, we reproduced the channel flow simulations presented in Ref.\ \cite{lamura_multi-particle_2001} for $\MM=35, \alpha=90$ degrees and $T^\ast=0.4$. We verified that the no-slip condition is satisfied. Moreover, from a fit of the velocity profile to the theoretical result for Poiseuille flow, 
\begin{equation} \label{parabola.eq}
	v_x(y) = \frac{f_{\rm ext}}{2\nu}y(L-y),
\end{equation}
we extract the fluid viscosity $\nu^\ast\approx 0.089\pm0.001$, in agreement with the findings reported in Ref.\ \cite{lamura_multi-particle_2001}. 

Having tested our numerical implementation in the uncoupled case, we now proceed to investigate the fully coupled system including angular momentum conservation and backflow effects ($n^\ast>0$). 
For the channel geometry shown in Fig.\ \ref{channel.fig}, we choose the channel width $L=32$ or $L=64$, length $50$, and $\MM=100$ particles per collision cell, i.e.\ in total $1.6$ or $3.2\times 10^5$ particles. 
For $T^\ast=0.1$ and no externally applied field, $h_0=0$, we find a parabolic velocity profile, from which we extract the kinematic viscosity of the base fluid from Eq.\ \eqref{parabola.eq} as $\nu_{\rm s}^\ast=0.114\pm0.001$. We refer to this as the solvent viscosity since the magnetic force density \eqref{Fmag} vanishes in this case. 
For the same conditions but in the quiescent state, we extract a self-diffusion coefficient $D\approx 0.05$, in agreement with the $\MM\gg 1$ limit of the theoretical expression, $D^\ast\to T^\ast \Delta t/2$. 
Therefore, the Schmidt number measuring the ratio of viscous over molecular diffusion becomes ${\rm Sc}=\nu_{\rm s}/D\approx 2.3$. 
The condition ${\rm Sc}>1$ indicates that collisional transport dominates over kinetic transport, which is the relevant regime for fluid simulations \cite{ripoll_dynamic_2005}. 
Particle-based flow simulations in general and MPC in particular do not strictly observe the incompressibility condition $\bnabla\cdot\bv=0$. 
For small Mach numbers,  
${\rm Ma} = v_{\rm max}/c_{\rm s}\ll 1$, incompressibility is approximately restored. 
Here, $v_{\rm max}$ is the maximal fluid velocity and $c_{\rm s}=\sqrt{5\kT/(3m)}$ the speed of sound. 
For the present conditions, $T^\ast=0.1, \MM=100$, and $f_{\rm ext}=5\times 10^{-5}$, we find $v_{\rm max}\approx 0.056$, leading to relatively small Mach numbers, ${\rm Ma}\approx 0.14$. Therefore, we can consider the fluid to be approximately incompressible. 
The Reynolds number is defined as ${\rm Re}=UL/\nu_{\rm s}$, where $U$ denotes a characteristic flow velocity. For $U=v_{\rm max}$, we obtain 
${\rm Re}\approx 15.7$, well in the laminar regime for channel flow. 
The Weissenberg number ${\rm Wi}=\tauB U/L$ quantifies the ratio of elastic to viscous forces. Choosing again $v_{\rm max}$ as the characteristic flow velocity $U$, we find ${\rm Wi}\approx 0.18$, which can be considered to be still in the Newtonian limit 
\cite{Ilg:mdferro04}. 

Having ensured that the simulations are performed in the proper parameter regime for laminar, viscous, near-incompressible flow, we now investigate the effect of an external magnetic field on the flow behavior. 
For simplicity, we choose $\chiL=0$, 
\changed{in agreement with the model \eqref{torque-balance},} 
so that demagnetization effects are absent and the internal field ${\bh}$ coincides with the external field ${\bh}_0$. 
Other parameters are chosen as $\tauB^\ast=100$ and $n^\ast=10^{-3}$.

\begin{figure}[hbt]
  \includegraphics[width=0.45\textwidth]{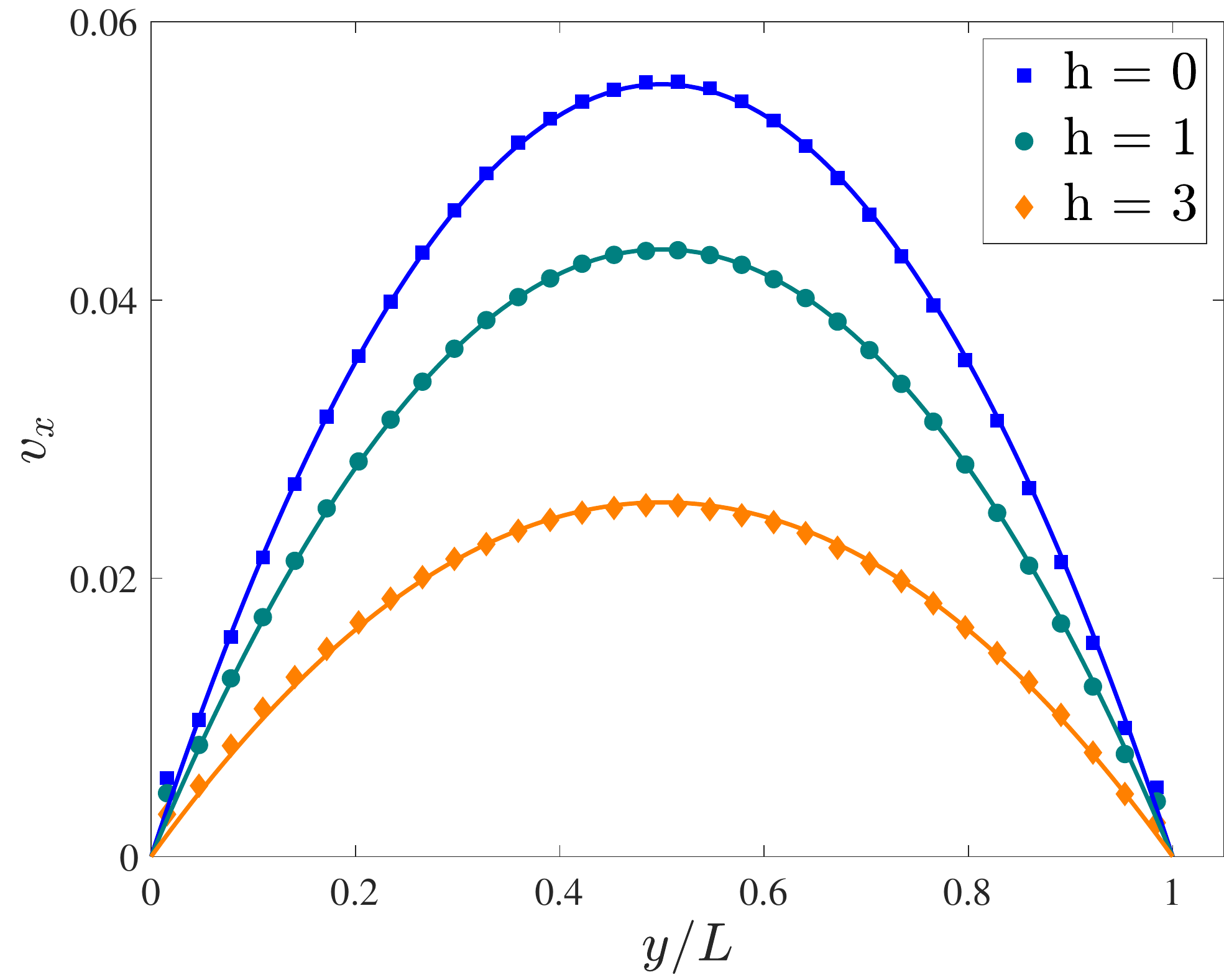}
  \caption{Velocity profiles for $f^\ast_{\rm ext}=5\times 10^{-5}, n^\ast=10^{-3}$, $\chiL=0$ and various strengths of the magnetic field $h$. Solid lines show fits to the parabolic profile, Eq.\ \eqref{parabola.eq}. }
  \label{velprofiles.fig}
\end{figure}

Figure \ref{velprofiles.fig} shows the flow profile for identical external forcing $f^\ast_{\rm ext}=5\times 10^{-5}$ but different strengths of the magnetic field $h$ applied in the gradient direction. 
We observe from Fig.\ \ref{velprofiles.fig} that the maximum flow velocity decreases with increasing magnetic field strength. 
We find that the Poiseuille profile \eqref{parabola.eq} fits the velocity field very accurately. From these fits, we extract a field-dependent effective kinematic viscosity $\nu(h)$. 
For $h=0$, we obtain the solvent viscosity, $\nu(0)=\nu_{\rm s}$, as discussed above. 
For $h>0$, we find that the effective viscosity increases with increasing magnetic field strength. This phenomenon was pioneered by McTague \cite{mctague} and is known in the literature as ``magnetoviscous effect'' \cite{ODENB02,Ilg_lnp}. 

\begin{figure}[hbt]
  \includegraphics[width=0.45\textwidth]{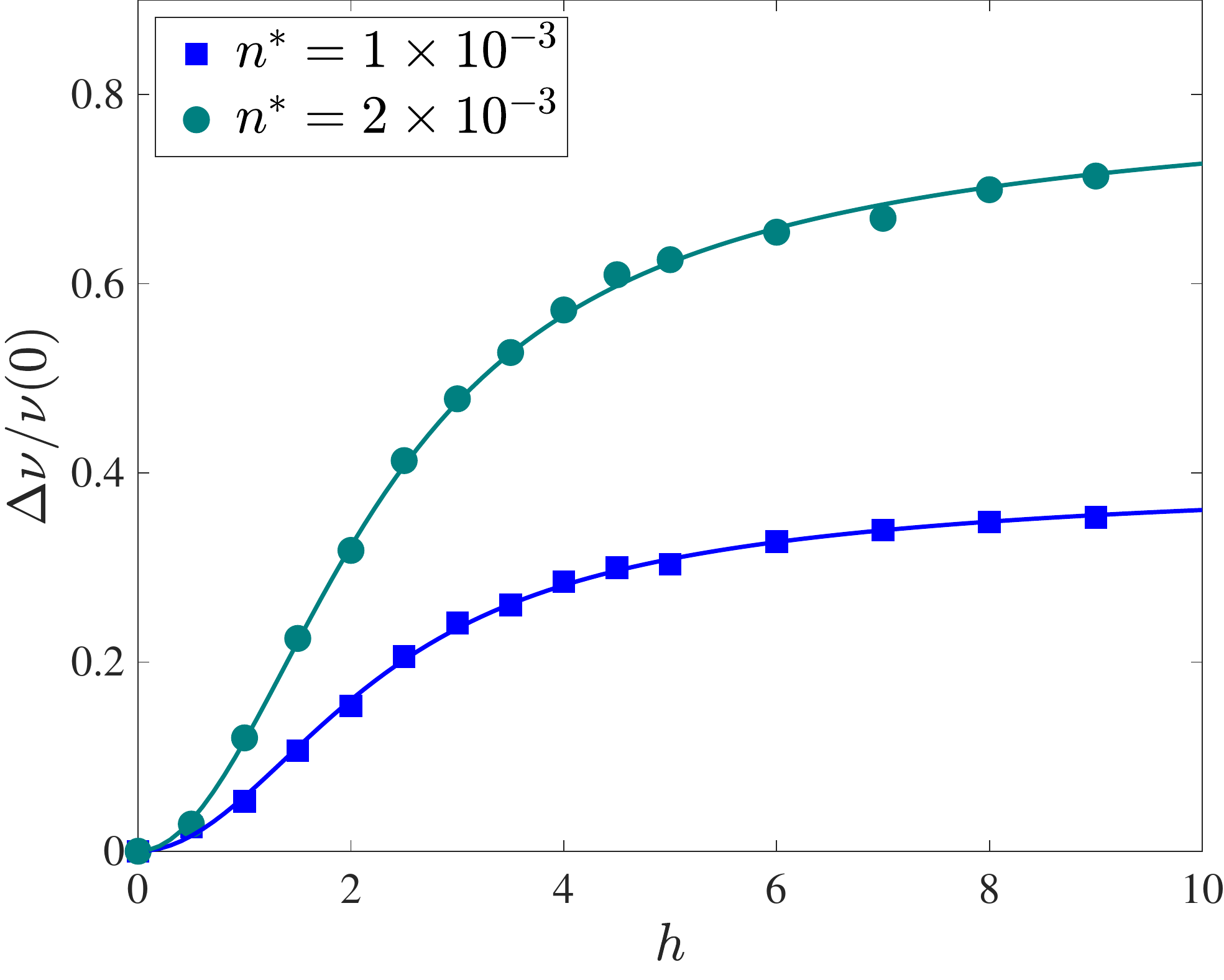}
  \caption{Dependence of the relative viscosity $\Delta\nu/\nu(0)$ on the dimensionless magnetic field strength $h$ for $n^\ast=10^{-3}$ and $n^{\ast}=2\times 10^{-3}$. Other parameters are chosen as $f^\ast_{\rm ext}=5\times 10^{-5}$, $\chiL=0$. Symbols show simulation results, whereas solid lines correspond to the theoretical result, Eq.\ \eqref{etarot.eq}.}
  \label{nueff-h.fig}
\end{figure}

Figure \ref{nueff-h.fig} shows the relative viscosity increase $\Delta\nu(h)/\nu(0)$, where $\Delta\nu(h) = \nu(h) - \nu(0)$ denotes the viscosity change and $\nu(h)$ is obtained from fits to the velocity profile  \eqref{parabola.eq} as described above. We verified that we obtain identical results within numerical accuracy for weaker forcing, $f^\ast_{\rm ext}=2\times 10^{-5}$. 
For weak flows, ${\rm Wi}\ll 1$, the model \eqref{torque-balance} can be solved analytically to give 
\begin{equation} \label{etarot.eq}
\frac{\Delta\nu(h)}{\nu(0)}=\frac{3}{2}\phi \frac{hL_1^2(h)}{h-L_1(h)},	
\end{equation}
where $L_1(h)=\coth(h)-1/h$ denotes the Langevin function \cite{MRS74,IKH01}. 
As can be seen from Fig \ref{nueff-h.fig}, the numerical simulations are in very good agreement with this theoretical result. 
Also the prefactor $(3/2)\phi$, giving the maximum of the relative viscosity increase, determined from fits to the numerical data ($\phi\approx 0.27\pm0.01$ for $n^\ast=10^{-3}$) are in good agreement with the relation $\phi=n^\ast\tauB^\ast/(3\nu_s^\ast)$ obtained above (giving here $\phi\approx 0.29$). 
Before commenting below on the slight deviation in the value of $\phi$, we want to mention that ferrofluids are typically dilute, corresponding to smaller values of $\phi$. For the purpose of demonstrating the method, however, the current choice of parameters should be sufficient to validate the correct implementation of the hybrid model combining MPC with the stochastic magnetization dynamics.  

Upon closer inspection, we find from the velocity profiles that the no-slip boundary condition and therefore the parabolic velocity profile is not perfectly satisfied. This effect is known in the literature and methods have been proposed to deal with this effect  \cite{JonathanKWhitmer:2010kp}. 

\begin{figure}[hbt]
  \includegraphics[width=0.47\textwidth]{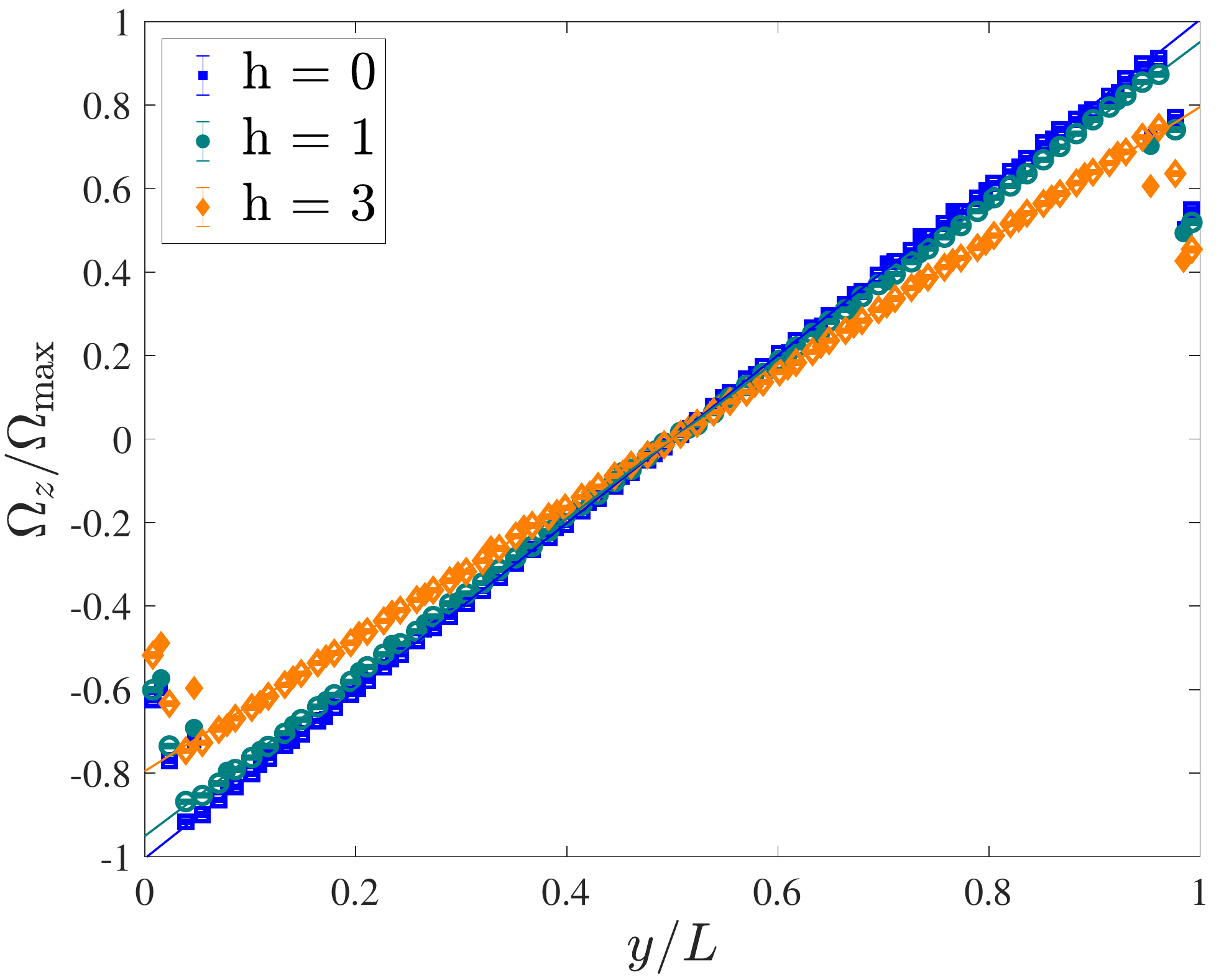}
   \includegraphics[width=0.47\textwidth]{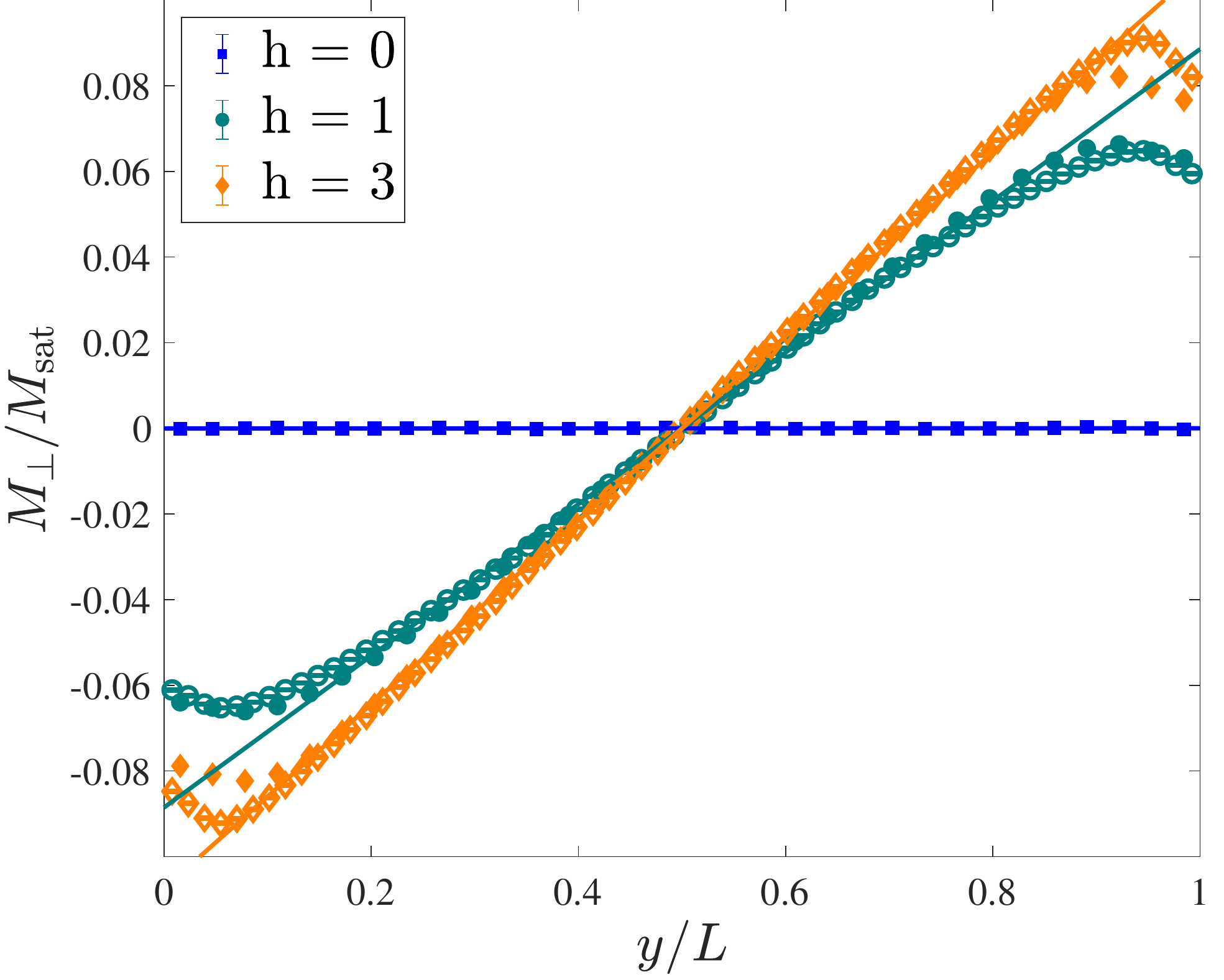}
 \caption{(a) The profile of the flow vorticity $\Omega_z/\Omega_{\rm max}$ is shown across the channel for the same conditions as in Fig.\ \ref{velprofiles.fig}. The position $y$ within the channel is scaled with the channel width $L$ and $\Omega_{\rm max}$ the maximum vorticity for $h=0$. Solid symbols correspond to $L=32$, open symbols to $L=64$. Solid lines show a linear fit. 
 (b) The perpendicular magnetization component $M_\perp$ normalized by the saturation magnetization $M_{\rm sat}$ is shown across the channel with $y$ for the same conditions as in Fig.\ \ref{velprofiles.fig}.}
  \label{Omega.fig}
\end{figure}

To investigate these deviations in more detail, we show in Fig.\ \ref{Omega.fig}(a) the profile of the flow vorticity $\Omega_z=(1/2)[\bnabla\times\bv]_z=-(1/2)\dd v_x(y)/\dd y$. The same conditions as used for Fig.\ \ref{velprofiles.fig} are chosen. We scale $\Omega_z$ with the maximum vorticity expected for Poiseuille flow in the field-free case, $\Omega_{\rm max}=f_{\rm ext}L/(4\nu_{{\rm s}})$. 
While the profiles are nicely linear in the center of the channel as expected for Poiseuille flow, deviations near the channel walls are apparent. 
As expected, these deviations are more pronounced in the vorticity compared to the velocity profile due to spatial gradients. 
We performed additional simulations for wider channels ($L=64$) and verified that the perturbation of the profile does not grow with $L$ but remains a boundary effect, propagating only a distance around $2a$ into the channel.  

Figure \ref{Omega.fig}(b) shows the perpendicular magnetization profile $M_\perp/M_{\rm sat}=\ave{u_y}$ across the channel. 
As expected, $M_\perp$ vanishes in the center of the channel since the vorticity is zero there. Overall, we find a linear profile of $M_\perp$ that follows the linear profile of the vorticity of the Poiseuille flow shown in Fig.\ \ref{Omega.fig}(a). However, deviations from the linear profile are apparent near the channel walls which result from deviations in the vorticity near the walls observed in Fig.\ \ref{Omega.fig}(a). 
It is worth noting that deviations from the linear magnetization profile occur over a slightly broader boundary layer compared to the deviations from the Poiseuille velocity profile. 

Discussing these boundary effects in more detail is beyond the scope of the present work. Here, we just want to mention that the reduced vorticity near the wall leads to a corresponding reduction of the local perpendicular magnetization component. As a consequence, the overall viscosity change $\Delta\nu$ is slightly reduced compared to the theoretical value, as seen in the slightly reduced value of $\phi$ found above.

\section{Results for flow around cylinder} \label{results-cylinder.sec}

As a further demonstration of the flexibility of the MPC method, we here consider 
the two-dimensional flow of a ferrofluid around a square cylinder of diameter $D$ inside a planar channel of width $L$ and length $L_x$. 
No-slip boundary conditions are imposed on the channel and cylinder walls by the same bounceback algorithm as used in Sect.\ \ref{results-channel.sec}. 
Following Ref.\ \cite{lamura_multi-particle_2001}, we impose flow by prescribing the average velocity 
$v_x(y)=\frac{4v_{\rm max}}{L^2}y(L-y)$ for particles within the inlet region $0\leq x <10$. In addition, periodic boundary in the flow direction are imposed. 
The horizontal position of the cylinder center is chosen as $L_x/4$, which was found to sufficiently reduce the influence of inflow and outflow boundary conditions \cite{breuer_accurate_2000}. 
Following previous studies \cite{lamura_multi-particle_2001,breuer_accurate_2000}, we choose the blockage ratio $D/L=1/8$.

Determining the flow around a cylinder is a classical problem in fluid dynamics where different flow regimes can be distinguished according to the value of the Reynolds number ${\rm Re}=v_{\rm max}D/\nu$. 
The creeping flow regime for ${\rm Re}<1$ is dominated by viscous forces and no separation is observed. 
For larger Reynolds numbers, the flow field separates at the downstream side of the cylinder, forming two steady, counter-rotating vortices behind the cylinder. The size of this recirculation region $L_{\rm r}$ increases with ${\rm Re}$ until the onset of the van-K\'arm\'an vortex street at a critical Reynolds number \cite{breuer_accurate_2000}. 

We here focus on the regime  $5 \leq {\rm Re} \leq 30$ where we expect to see steady recirculation regimes.  
In order to obtain reasonable spatial resolution, we choose $L=200, L_x=600$. 
For the average number of particles per collision cell, we set $Q=30$, which means the simulations contain around $3.6\times 10^6$ particles.  
The temperature is chosen as $T^\ast=0.05$, corresponding to a viscosity $\nu_{\rm s}^\ast\approx 0.0913\pm 0.0001$. 
Furthermore, to observe magnetic field-induced changes in the velocity field more clearly, we extend the range of concentrations to 
$n^\ast = 5\times 10^{-3}$. 
Choosing again $\chiL=0$ eliminates contributions of the demagnetization field. 
Simulations are performed for a total of $10^5$ integration steps and averages extracted after $5\times 10^4$ steps. 
We calculate two-dimensional velocity and magnetization fields and their derivatives using kernel smoothing methods as described in the appendix 
\ref{kernelsmooth.sec}. 

\begin{figure}[hbt]
  \includegraphics[width=0.48\textwidth]{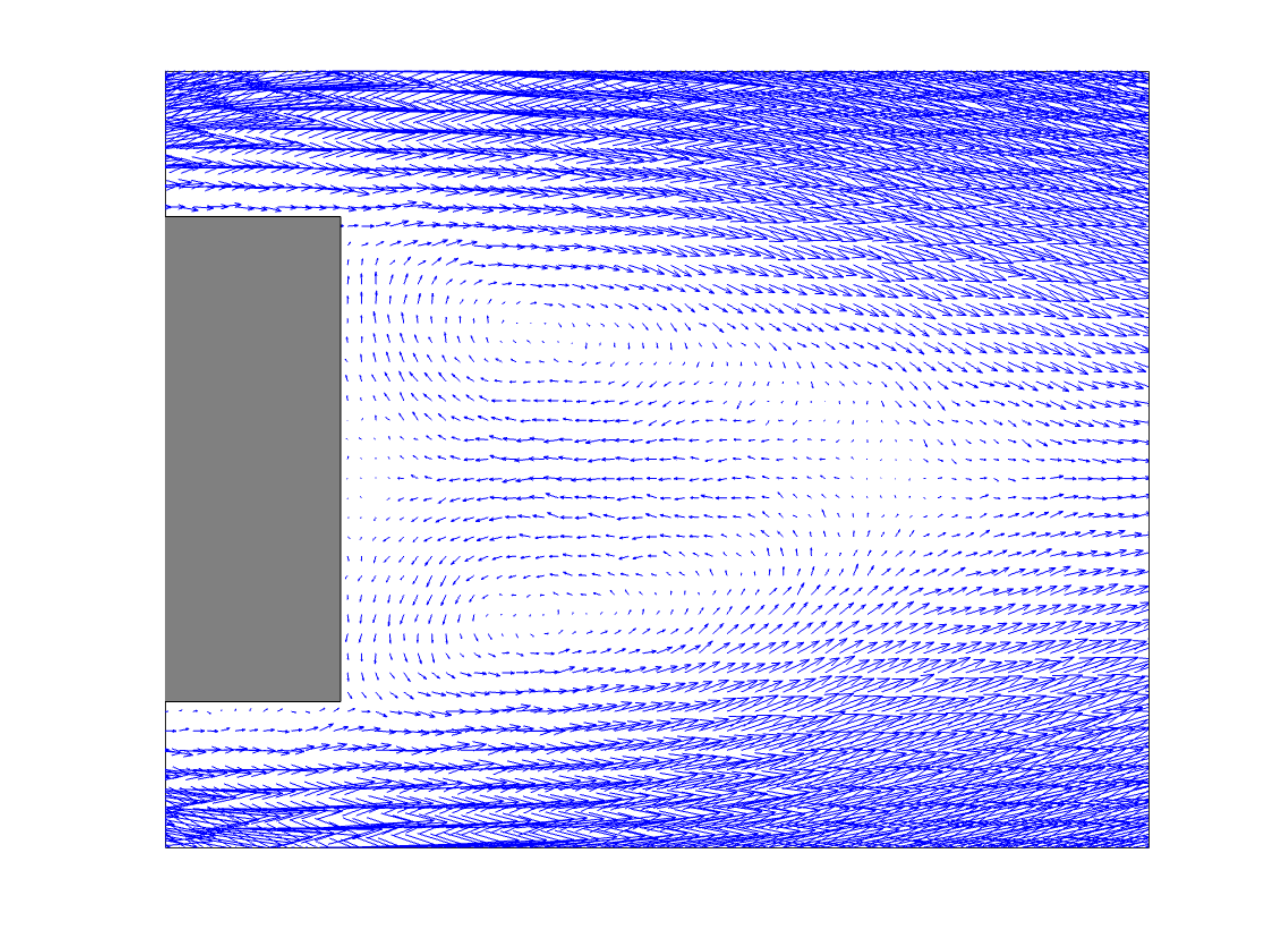}
  \includegraphics[width=0.48\textwidth]{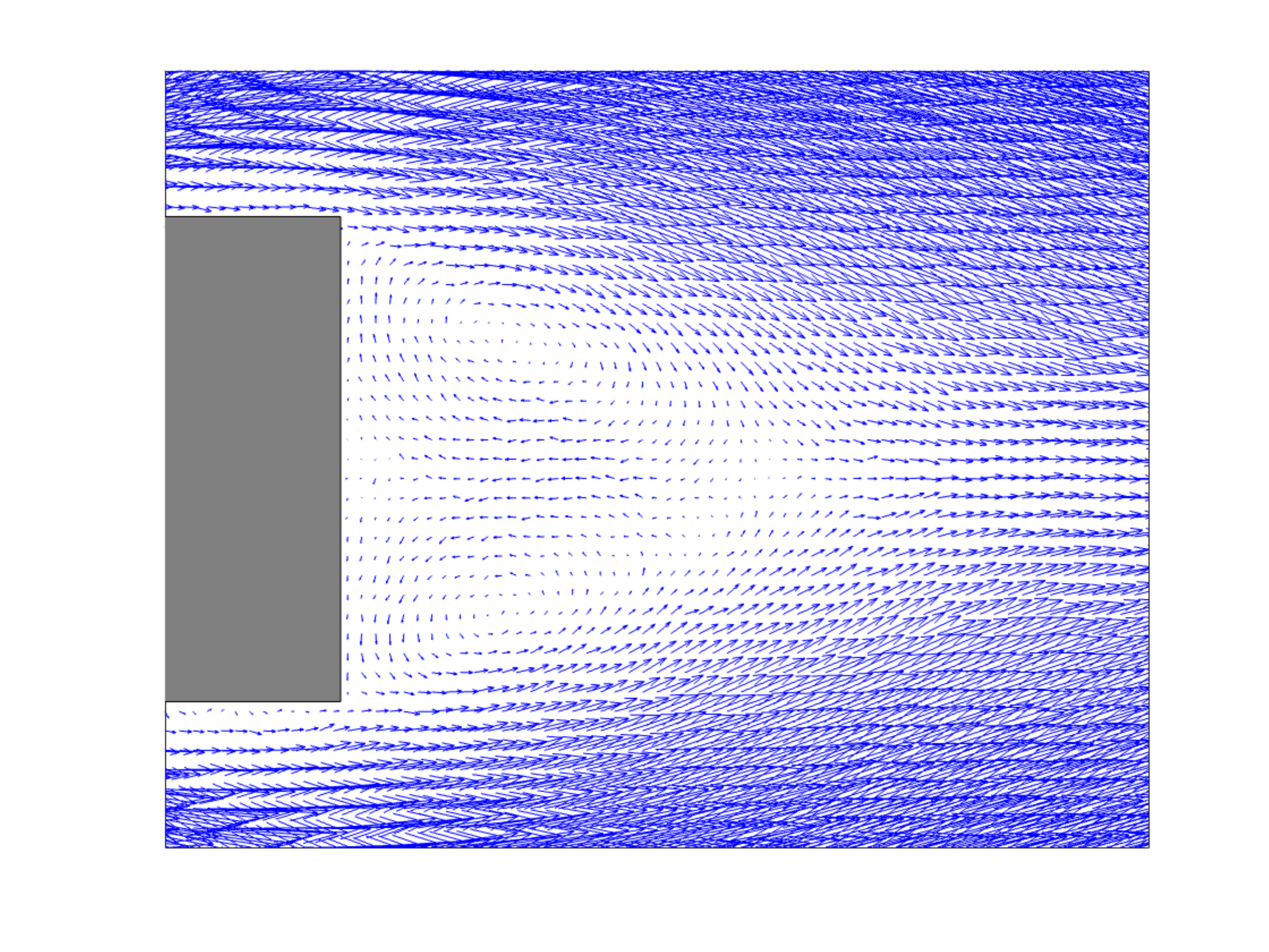}
  \caption{Stationary velocity field for the flow around a square cylinder (gray) for $h=0$ (top) and $h=4$ (bottom). 
  For better visibility, only a part of the velocity field is shown. The simulation parameters are chosen as $T^\ast=0.05$, $n^\ast=5\times10^{-3}$, $\chiL=0$, $v_{\rm max}=0.1$.}
  \label{Vfield.fig}
\end{figure}

\begin{figure}[hbt]
  \includegraphics[width=0.48\textwidth]{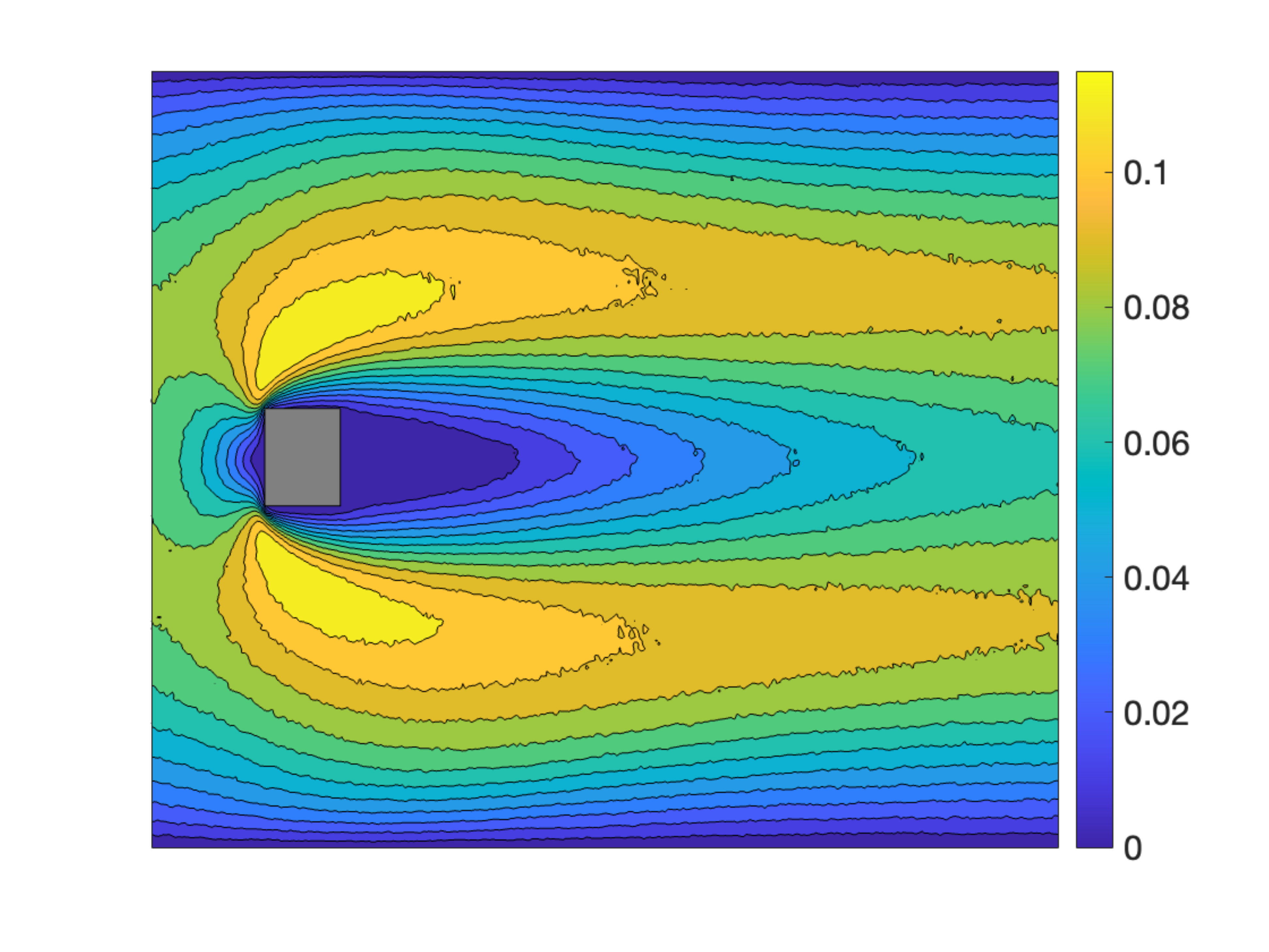}
  \includegraphics[width=0.48\textwidth]{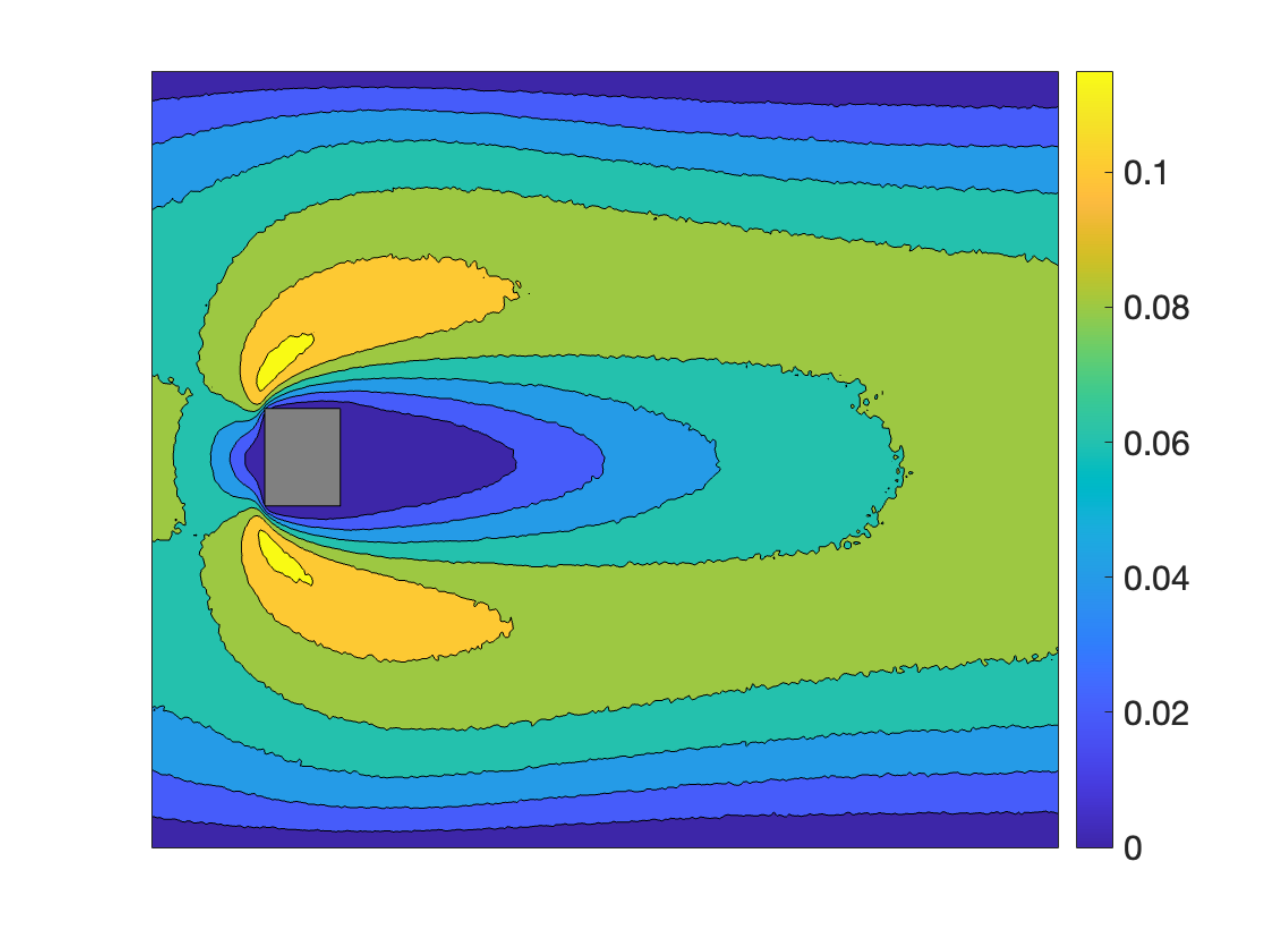}
  \caption{The color-coded magnitude of the stationary velocity field is shown ($h=0$ in the top and $h=4$ in the bottom panel) for the same conditions as in Fig.\ \ref{Vfield.fig}, but for a larger part of the flow field.}
  \label{Vmag.fig}
\end{figure}

Figure \ref{Vfield.fig} shows the stationary velocity fields with and without applied magnetic field for a particular choice of model parameters 
($T^\ast=0.05, n^\ast=5\times10^{-3},v_{\rm max}=0.1$). 
We observe that the applied magnetic field changes the flow field and causes a smaller recirculation region. 
Qualitatively, such a change is expected due to the field-induced increase in viscosity. 
Looking at a larger portion of the velocity field, Fig.\ \ref{Vmag.fig} shows the significant changes the external magnetic field induces not only on the wake, but also on the larger-scale flow field.

\begin{figure}[hbt]
  \includegraphics[width=0.48\textwidth]{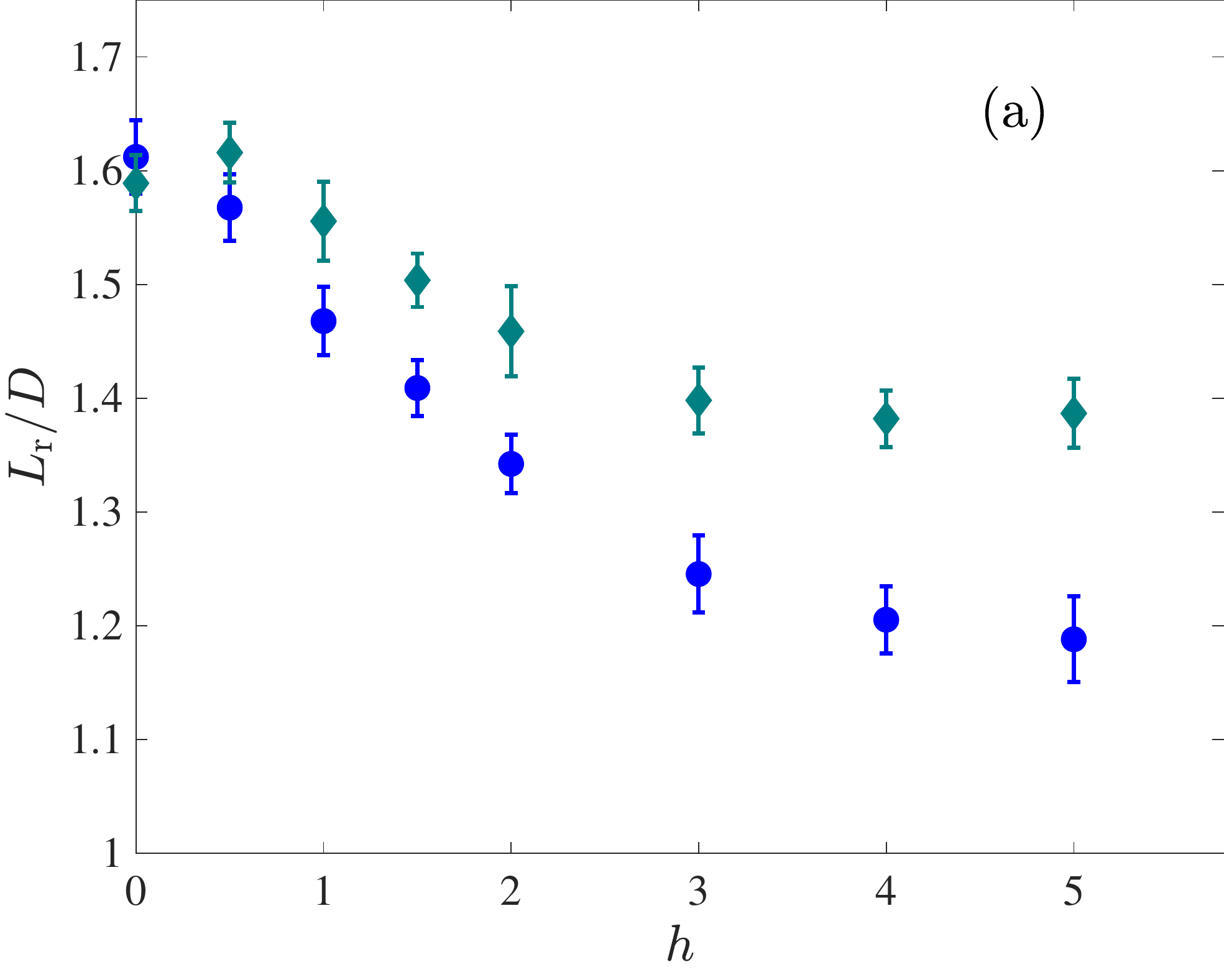}
  \includegraphics[width=0.48\textwidth]{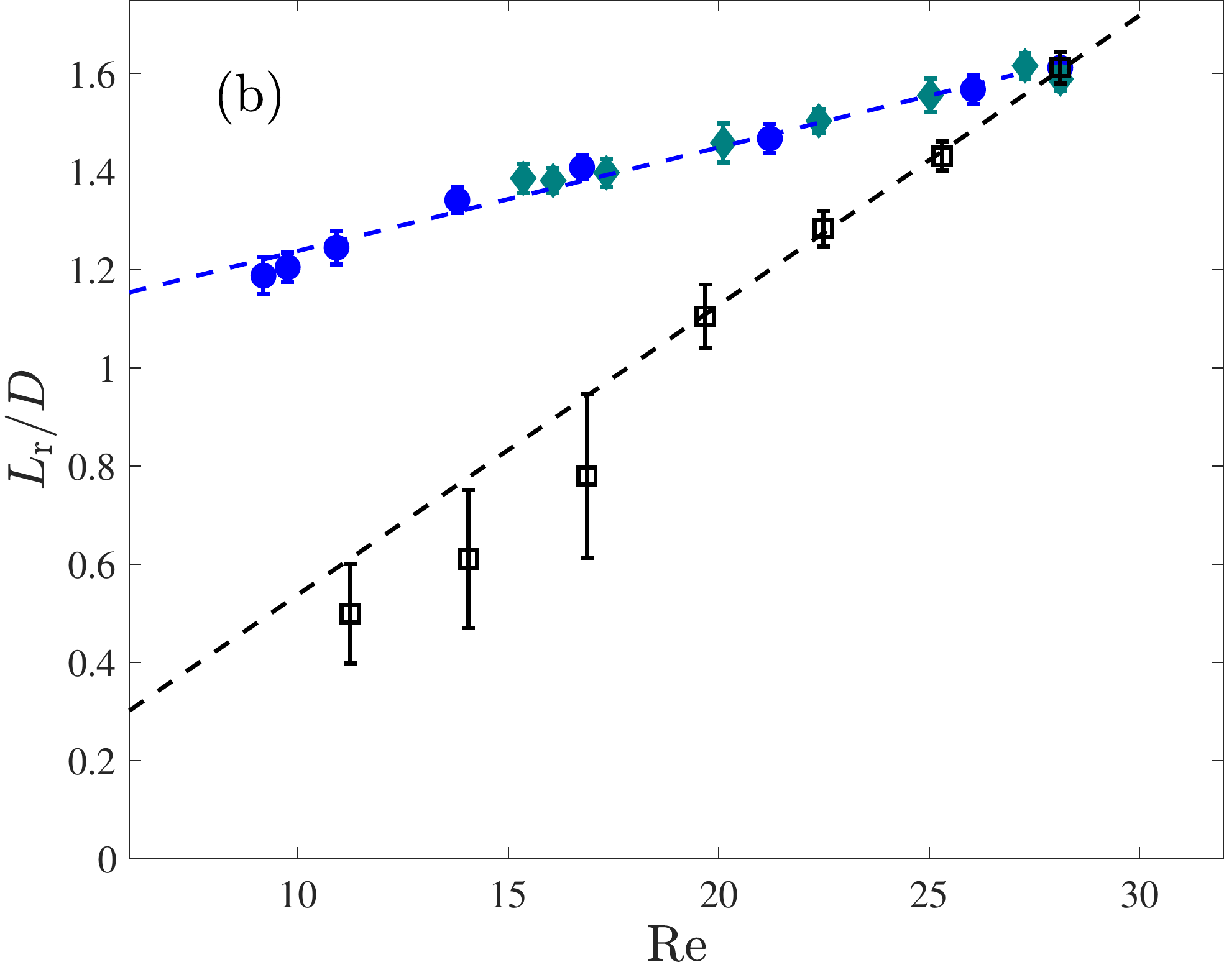}
  \caption{(a) Length of the recirculation region $L_{\rm r}$ scaled with the diameter of the cylinder $D$ as a function of the magnetic field strength $h$. Diamonds and circles correspond to $n^\ast=2\times 10^{-3}$ and $5\times10^{-3}$, respectively.  Other parameters are chosen as $T^\ast=0.05, v_{\rm max}=0.1$. 
  (b) The data in panel (a) are shown as a function of effective Reynolds number. In addition, open black squares show the result for $h=0$, $n^\ast=5\times10^{-3}$ but varying $v_{\rm max}$.}
  \label{Lr.fig}
\end{figure}

To make these observations more quantitative, we extract the length of the recirculation zone, $L_{\rm r}$, from the velocity fields. 
In particular, we determine $L_{\rm r}$ from the zero-crossing of the $x$-component of the centerline velocity at the end of the wake. 
We use a parabolic fit to the centerline velocity profile near the end of the wake to determine $L_{\rm r}$ and estimate error bars based on uncertainties in the fit parameters. 
Figure \ref{Lr.fig} shows the length of the recirculation region $L_{\rm r}$ in units of the diameter $D$ of the cylinder as a function of the magnetic field and the Reynolds number. 
Increasing the magnetic field strength leads to a decrease of $L_{\rm r}$, the effect being more pronounced for larger concentrations. 
This observation is consistent with the increased effective viscosity we found in Sect.~\ref{results-channel.sec}. 
In the absence of an external magnetic field, by varying the maximum inflow velocity $v_{\rm max}$, we recover earlier results showing 
a linear increase of the length of the recirculation zone with Reynolds number, 
\begin{equation} \label{LrproptoRe}
L_{\rm r}/D = c_0 + c_1\ {\rm Re}
\end{equation}
with $c_1 \approx 0.059 \pm 0.007$, indicated as black dashed line in Fig.~\ref{Lr.fig}(b). 
Keeping instead $v_{\rm max}=0.1$ fixed but varying the magnetic field strength $h$, we also find the linear relationship Eq.\ \eqref{LrproptoRe} 
when plotting the data in Fig.~\ref{Lr.fig}(a) against the effective Reynolds number ${\rm Re}=v_{\rm max}D/\nu(h)$, with $\nu(h)$ the field-enhanced effective kinematic viscosity discussed in Sect.~\ref{results-channel.sec}. 
Interestingly, however, we observe that $L_{\rm r}$ varies much less strongly with ${\rm Re}$ due to a magnetic field compared to the non-magnetic case. 
Indeed, we find there a slope $c_1 \approx 0.021 \pm 0.003$, less than half the value  in the field-free case when varying $v_{\rm max}$. 

\begin{figure}[hbt]
  \includegraphics[width=0.48\textwidth]{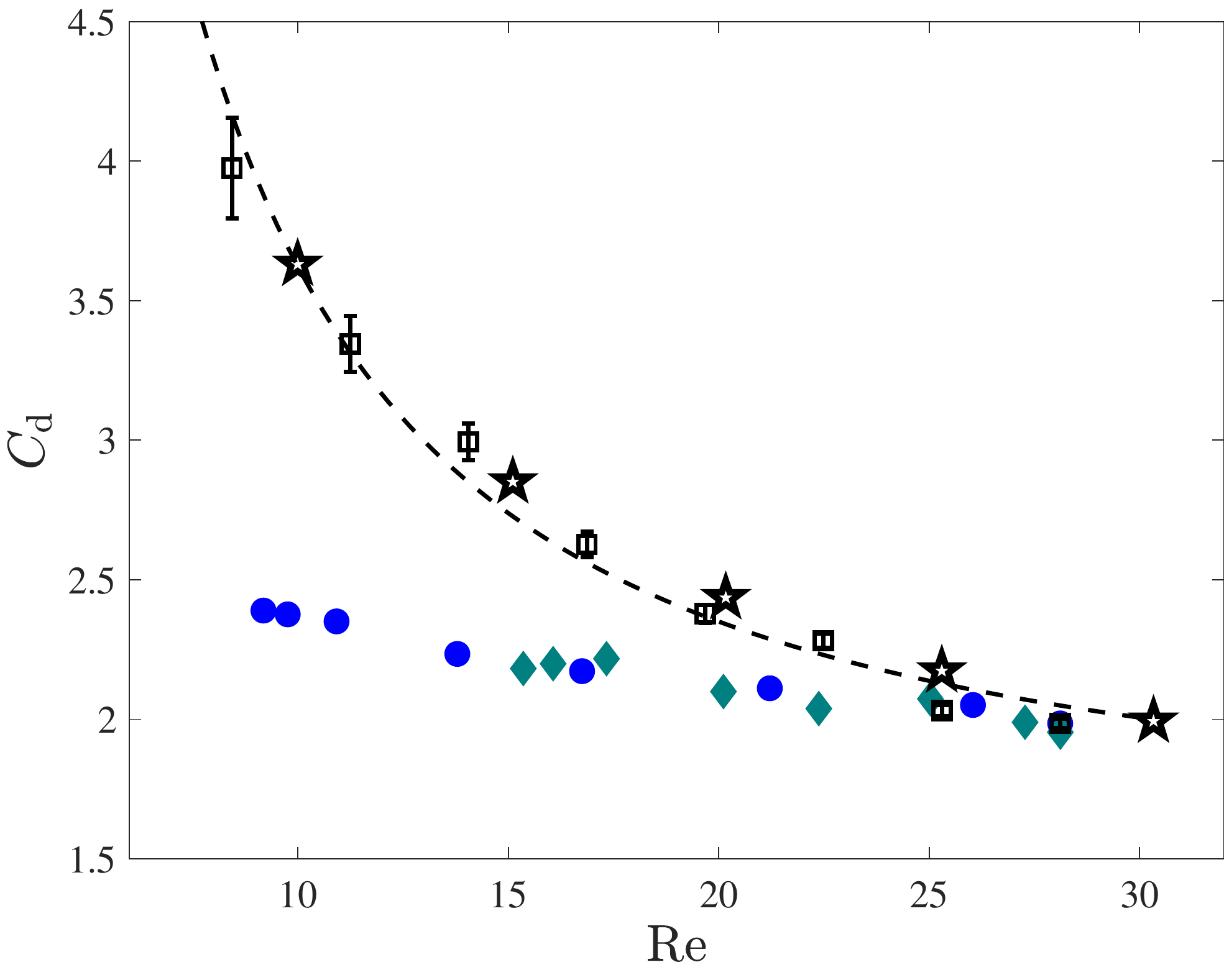}
  \caption{The dimensionless drag coefficient $C_{\rm d}$ is shown as a function of Reynolds number for the same conditions and color coding as in Fig.~\ref{Lr.fig}. Stars indicate the results of Ref.~\cite{breuer_accurate_2000} and dashed-line a power-law fit.}
  \label{Cd.fig}
\end{figure}

To further illustrate the effect of a magnetic field on ferrofluid flow, we also study the drag coefficient 
$C_{\rm d}$. The drag coefficient is defined in two dimensions as 
$C_{\rm d}=2F_{\|}/(Qv_{\rm max}^2D)$, where $F_{\rm \|}$ denotes the force in flow direction exerted by the fluid on the cylinder \cite{breuer_accurate_2000,lamura_multi-particle_2001}. 
Within the MPC scheme, the force $F_{\|}$ can readily be evaluated from the parallel component of the total momentum the particles transfer to the cylinder wall when bouncing back. 
Figure \ref{Cd.fig} shows the drag coefficient $C_{\rm d}$ as a function of Reynolds number. In the absence of an externally applied field (black open symbols), we observe the characteristic strong decrease of $C_{\rm d}$ with ${\rm Re}$, approximately as an inverse power law, $C_{\rm d}\sim{\rm Re}^{-z}$ with $z\approx1.35\pm0.15$, in agreement with finite-volume simulations in Ref.~\cite{breuer_accurate_2000}. 
However, when a magnetic field is applied (filled colored symbols), we observe that the drag coefficient is significantly smaller compared to the non-magnetic fluid at the same Reynolds number. 
We observe this magnetic drag reduction in the low-Reynolds number regime $5\lesssim {\rm Re}\lesssim20$. 

\section{Conclusions} \label{conclude.sec}

We here present an extension of the MPC method to simulate dynamics and flow of magnetic fluids, including fluctuation and backflow effects. 
In analogy to recent extensions of MPC to nematic liquid crystals \cite{Lee:2015iv,Mandal:2019bs}, 
we equip each MPC particle with a magnetic moment and include stochastic magnetization dynamics via additional rotational motion of the individual particles. 
Fluid and magnetization dynamics are coupled to each other via velocity gradients and the magnetic force density. 
We successfully tested this hybrid scheme in a standard, two-dimensional channel geometry. For Poiseuille flow, we reproduce the magnetoviscous effect and recover quantitatively the theoretical result for the relative viscosity increase with magnetic field strength. 
Using standard methods to implement no-slip conditions in MPC simulations, we nevertheless observe small deviations of the velocity gradient from the theoretical profile very close to the wall, leading to corresponding deviations in the magnetization profile. This effect is already known in the literature and more refined methods have been proposed to better realize the no-slip condition \cite{JonathanKWhitmer:2010kp}. 
We leave this more technical aspect for future research. 
We also illustrate the new MPC method for magnetic fluids for the benchmark problem of flow around a square cylinder. 
In the absence of an applied field, we verify known results for the length of the recirculation region behind the cylinder increasing linearly with Reynolds number and the decrease of the drag coefficient for non-magnetic fluids. 
In addition, we study the dependence of these quantities on an externally magnetic field for ferrofluids. 
We observe that the length of the wake is increased and the drag is reduced by a magnetic field compared to the non-magnetic case at the same effective Reynolds number. 
Being able to reduce drag by a magnetic field in the low-Reynolds number regime might be of interest in microfluidics applications. 
More generally, we find that these quantities are not described by the effective Reynolds number alone, even when accounting for the field-induced effective viscosity. This is likely due to the anisotropy of the effective viscosity induced by the magnetic field  \cite{mctague,Sreekumari2015}, which leads to ferrofluids showing different flow behavior than ordinary viscous fluids. 

The method proposed here to simulate ferrofluid flow naturally includes thermal fluctuations that are particularly relevant at small scales. 
Furthermore, \changed{as a solver for fluctuating ferrohydrodynamics} the method inherits all the benefits of the MPC approach, which can easily be extended to three spatial dimension and more complicated geometries. 
The MPC method is also particularly well-suited as an efficient way to model solvent effects in colloidal suspension \cite{malevanets_solute_2000}. 
Future studies may also include the effect of demagnetization field that we have neglected here. 
In addition, the method is very flexible and can straightforwardly accommodate different magnetization equations, \changed{such as e.g.\ the so-called chain model \cite{ZI00} to describe chain-forming ferrofluids or including mean-field interactions \cite{Ilg_lnp}}. 
Therefore, we expect this method to be a useful tool for the simulation of \changed{flow and} hydrodynamic effects in magnetic fluids.

\section*{Acknowledgments}
Discussions with Anoop Varghese during a very early stage of this project are gratefully acknowledged. 

\begin{appendix}

\section{Stochastic Heun algorithm for orientations} \label{Heun.sec}

The stochastic Heun algorithm is a predictor-corrector scheme with second-order weak convergence in the Stratonovich sense \cite{GarciaPalacios:2000}. 
For rotational motion, we need to ensure that the orientations $\bu_i(t)$ remain unit vectors for all times. 

In the predictor step, new orientations $\breve{\bu}_i$ are calculated from an Euler scheme as indicated in Sect.\ \ref{model.sec}, 
\begin{align}
	\breve{\bu}_i & = \frac{\bu_i^{\rm P}}{|\bu_i^{\rm P}|},\\
	\bu_i^{\rm P} & = \bu_i(t) + \Delta\bomega_i(t)\times\bu_i(t)
\end{align}
and $\Delta\bomega_i(t)$ is given by Eq.\ \eqref{delta-omega} evaluated at time $t$. 

These predicted orientations $\breve{\bu}_i$ are now used to computed new angular velocity changes $\Delta\breve{\bomega}_i$, where for the latter, the right hand side of Eq.\ \eqref{delta-omega} is evaluated with $\bu_i^{\rm P}$ instead of $\bu_i(t)$. 

In the corrector step, the new orientations are obtained from 
\begin{align}
	\bu_i(t+\Delta t) & = \frac{\bu_i^{\rm C}}{|\bu_i^{\rm C}|},\\
	\bu_i^{\rm C} & = \bu_i(t) + \frac{1}{2}[\Delta\bomega_i(t)\times\bu_i(t) + \Delta\breve{\bomega}_i\times\breve{\bu}_i]
\end{align}

\section{Kernel smoothing} \label{kernelsmooth.sec}
Since evaluation of velocity and magnetostatic fields are important for the model, we give here some details on their evaluation in the simulation. 

We exemplify the method for the velocity component in flow direction, $v_x$. The other fields are evaluated in the same manner. 
First, we use the Nadaraya-Watson kernel regression estimator \cite{Loader} 
to find the instantaneous field as 
\begin{equation} \label{kernelsmooth.eq}
v_x(\br;t) = \frac{\sum_{i=1}^{N}v_{x,i}(t)K\left( \frac{|\br_i(t)-\br|}{b} \right)}{\sum_{j=1}^{N}K\left( \frac{|\br_j(t)-\br|}{b} \right)}
\end{equation}
where $K(z)$ is known as the kernel and the parameter $b$ as bandwidth or smoothing length. 
While the uniform kernel is frequently used, we here employ the Epanechnikov kernel 
\begin{equation} \label{Epa.eq}
	K(z) = \left\{ \begin{array}{cc}
	\frac{3}{4}(1-z^2) & |z|\leq 1\\
	0 & {\rm else} 	
	\end{array}\right. 
\end{equation}
which is known to minimize the mean integrated square error. We found that choosing $b$ equal to the linear size of the collision cells $a$ provides a good compromise between smoothing and keeping local fluctuations. 

Since there is a systematic attenuation bias due to smoothing, 
we resort to finite-difference schemes to calculate spatial gradients.  
Having evaluated the instantaneous velocity field $v_x$ at the centers of the collision cells from Eq.\ \eqref{kernelsmooth.eq}, we use the central difference scheme to approximate the  spatial partial derivatives as 
\begin{equation} \label{dvdy}
	\frac{\partial v_x(\br;t)}{\partial y} \approx \frac{v_x(\br+a\hat{\bf k};t)-v_x(\br-a\hat{\bf k};t)}{2a}
\end{equation}
with the unit vector $\hat{\bf k}$ defined in Fig.\ \ref{channel.fig}. 
And correspondingly with the unit vector $\hat{\bf i}$ in the $x$-direction for the partial derivative with respect to $x$. 
Only near the channel walls we use a first-order approximation instead. 
We found that the central difference scheme provides good results, also in comparison to using higher-order, differentiable kernel functions.

\section{Magnetostatics in channel geometry}	 \label{magnetostatics.sec}
For non-conducting fluids, Maxwell's equations are given by Eqs.\ \eqref{magnetostatics.eq}
and a separate, medium-dependent magnetization equation. 
First, we consider the exterior of the system to be non-magnetic,  
\begin{equation}
\bM_{0} = 0, \bB_{0}=\mu_{0}\bH_{0}
\end{equation}
and we assume the fields $\bH_{0}$ (and consequently $\bB_{0}$) are spatially uniform. 
Thus, Maxwell's equation \eqref{magnetostatics.eq} are identically satisfied in the exterior, 
$\bnabla\times\bH_{0} = {\bf 0}$ and $\bnabla\cdot\bB_{0} = 0$. 

Let us denote the fields inside the fluid as $\bH_{\rm }$ and $\bB_{\rm }$ with 
\begin{align}
\bB_{\rm } & = \mu_{0}(\bH_{\rm } + \bM)
\end{align}
Define the contribution $\bH_{\rm ff}$ of the magnetic fluid to the internal field, 
\begin{equation}
\bH_{\rm } = \bH_{0} + \bH_{\rm ff}
\end{equation}
Note that $\bH_{\rm ff}$ is often denoted in terms of a ``demagnetization field'', $\bH_{\rm } = \bH_{0} - {\bf D}\cdot \bM$, 
with ${\bf D}$ the demagnetisation tensor, which depends on the shape of the sample. 
Note that ${\bf D}$ is symmetric, has trace one, and all diagonal elements are non-negative 
\cite{moskowitz_theoretical_1966}. 

We now specialize to the channel geometry sketched in Fig.\ \ref{channel.fig}. From the continuity condition on the magnetic field, Eq.\ \eqref{field-continuity.eq}, we find that 
\begin{align}
B_{{\rm }y} & = B_{0,y} \\
H_{{\rm }x} & = H_{0,x} \quad \Rightarrow H_{{\rm ff},x}=0
\end{align}
Thus, we know that $\bH_{\rm ff}=H_{\rm ff}\hat{\bf k}$ is oriented normal to the wall. 
Next, to satisfy Maxwell's equation $\bnabla\times\bH={\bf 0}$ we require $\bnabla\times\bH_{\rm ff}={\bf 0}$ and therefore conclude that $H_{\rm ff}=H_{\rm ff}(y)$ must be independent of the horizontal position $x$ along the channel. 
Similarly, we can conclude from $\bnabla\cdot\bB_{\rm }=0$ and the continuity condition that $\partial B_{{\rm }x}/\partial x=0$ and therefore $\partial M_x/\partial x=0$ since $B_{{\rm }y}$ is constant. Therefore, the magnetization component $M_x$ is also independent of the horizontal position along the channel. 
Finally, we can also write $\bnabla\cdot\bB_{\rm }=0$ as $\bnabla\cdot(\bH_{\rm }+\bM)=0$, which leads to $\partial(H_{\rm ff}+M_y)/\partial y=0$. Since we already concluded that $H_{\rm ff}$ is independent of $x$, so must $M_y$ to satisfy this condition for all points within the channel. Therefore, none of the magnetostatic fields depends on the $x$-position within the channel.

\end{appendix}


%

\end{document}